\documentclass[10pt]{article}
\usepackage{amsmath}
\usepackage{amssymb}
\usepackage{graphicx}

\def\aa{A\&A}

\def\mnras{MNRAS}

\begin{document}

\setcounter{figure}{0}
\setcounter{table}{0}
\setcounter{footnote}{0}
\setcounter{equation}{0}

\vspace*{0.5cm}

\noindent {\Large ON FUTURE OPPORTUNITIES TO OBSERVE GRAVITATIONAL SCATTERING OF MAIN BELT ASTEROIDS INTO NEO SOURCE REGIONS}
\vspace*{0.7cm}

\noindent\hspace*{1.5cm} A. IVANTSOV, S. EGGL, D. HESTROFFER, W. THUILLOT\\
\noindent\hspace*{1.5cm} Institut de M\'{e}canique C\'{e}leste et de Calcul des \'{E}ph\'{e}m\'{e}rides - Observatoire de Paris\\
\noindent\hspace*{1.5cm} Avenue Denfert-Rochereau 77, 75014 Paris, France\\
\noindent\hspace*{1.5cm} e-mail: ivantsov@imcce.fr\\

\vspace*{0.5cm}

\noindent {\large ABSTRACT.} Orbital resonances are believed to be responsible for the delivery of main belt asteroids to the inner Solar System. Several possibilities have been suggested to transport asteroids and their fragments into mean motion and secular resonances including non-gravitational forces and gravitational scattering. We investigate future resonance crossings of known asteroids that occur in the main belt over the next century. Our goal is to identify potentially observable injections of asteroids into source regions for Near Earth Objects (NEOs) as well as to determine the role of close encounters among main belt asteroids in this process.

\vspace*{1cm}

\noindent {\large 1. INTRODUCTION}

\smallskip

Physical and orbital properties of the current NEO population can be explained when one assumes that their primary origin lies in the asteroid main-belt and Jupiter-family comet (P$<$20 years) regions (Bottke et al., 2000a, 2002; Greenstreet et al., 2012; Mainzer et al., 2012). Orbital resonances with the Gas Giants within the main-belt which cause strong eccentricity-pumping are transport mechanisms that can carry main belt asteroids into the inner Solar System. Two well known resonances in this respect are the 2:1 and 3:1 mean-motion resonance with Jupiter as well as the $\nu_5$ and $\nu_6$ secular resonances. Since the average life-time of asteroids in resonances is much shorter than the age of the Solar System, a constant flux of asteroids into resonant regions is necessary in order to sustain the observed NEO population (Morbidelli et al., 2002). 

Non-gravitational forces such as the Yarkovsky effect can cause a slow drift into resonance (Bottke et al., 2000b). Also, close encounters among asteroids lead to orbital migration (Delisle \& Laskar, 2012; Carruba et al., 2012, 2013). Yet, the importance of gravitational interactions between asteroids with respect to further moving future NEOs towards resonances is not well understood. This is due to computational difficulties arising with the numerical solution of the gravitational N-body problem. Results of multiple scattering events, for instance, strongly depend on the number of interacting bodies considered. Using high precision orbit predictions over one hundred years, we identify main belt asteroids that are injected into - or stay in the vicinity of relevant resonant regions that might cause them to become NEOs in the far future. We, thus, intend to provide interesting targets for astrometric observation campaigns which may lead to a better understanding of the processes involved in the generation of NEOs.

\vspace*{0.7cm}

\noindent {\large 2. METHOD}

\smallskip

In a first step, we identified those asteroids that are likely candidates to be scattered into a resonance. For this purpose the analytically calculated proper element catalog from AstDys has been searched and asteroids that are close to secular and mean motion resonances have been identified. The mean motion resonances (MMR) considered are 3:1, 5:2, 2:1, and 7:3 between asteroids and Jupiter. Secular resonances were included up to second order in g and s (e.g. Machuca \& Carruba, 2012). Our current sample encompasses asteroids with a distance of no more than 0.5 [''/yr] from the secular resonant frequencies and a maximum proper semi-major axis distance of $10^{-3}$ au from the locations of the four MMRs that are considered to be source regions for NEOs.

In order to determine, whether any of the sampled asteroids will be pushed into or at least closer to a resonance within the next century, we performed an exploratory numerical propagation of 1273 asteroids that where closest to the investigated secular resonances, and 52 asteroids that were close to the previously mentioned MMRs. Perturbations from all planets (DE431) as well as from the biggest 16 asteroids were considered. The equations of motion include all relevant post-Newtonian terms, as well as the $J_2$ form factor of the Sun. Initial conditions of the main belt objects (MBOs) were acquired from the HORIZONS for consistency. The proper elements of asteroids were computed using OrbFit 4.2 (Milani \& Gronchi, 2010).

\vspace*{0.7cm}

\noindent {\large 3. PRELIMINARY RESULTS}

\smallskip

A comparison of analytical proper elements for all numbered asteroids once for J2000 and once a century later yields that, in fact, 18 out of the 52 near mean motion resonant asteroids stay in the vicinity of their respective MMRs over 100 years, and 4 of them might even become resonance crossers during this time. For nearly secular resonant asteroids, 455 out of 1273 remained close to their resonance and 128 potential resonance crossing events were recorded. A shortened list of possible resonance crossing asteroids is given in Table 1.

\begin{table}[h]
\begin{center}
\begin{tabular}{c | l}
\hline
MMR\\
\hline
7:3 & 271956, 299835, 332810, 338943\\
\hline
SECULAR\\
\hline
$\nu_6+\nu_{16}$ & 11094, 112561, 112588,  ...\\
2$\nu_5+\nu_6$ & 103395, 127993, 13106, 132964,...\\
2$\nu_6+\nu_5 $ & 103774, 104624, 195280, 321328, ...\\
2$\nu_6-\nu_5-\nu_{16}$ & 108497, 148758, 177861, 203561, ...\\
3$\nu_6 - \nu_5$ & 146121, 176517 \\
2$\nu_6+\nu_{16}$ & 6234, 47790, 146664, 90239,...\\
3$\nu_6-2\nu_5$ &54486, 367618 \\
$\nu_6+2\nu_{16}$ & 164537\\
2$\nu_6-\nu_5$ & 75399, 233930 \\
2$\nu_6+\nu_{16}$ & 24986, 252191, 234075\\ 
\hline
\end{tabular}
\caption{Identified asteroids that may experience resonance crossing over the next century.}
\end{center}
\end{table}

\vspace*{0.7cm}

\noindent {\large 4. CONCLUSIONS}

\smallskip
We have found considerable dynamical mixing close to investigated resonances since only roughly one-third of our asteroids' sample stayed close to the respective resonances within the next 100 years. About one-tenth of the investigated population was identified as potentially resonance crossing. A more detailed modeling of the gravitational scattering processes will provide information on the importance of asteroid-asteroid interaction around resonances relevant to NEO production.

\vspace*{0.7cm}

\noindent {\large 5. REFERENCES}
%
%
%
%
%
{

\leftskip=5mm
\parindent=-5mm

\smallskip

Bottke, W.~F., Jedicke, R., Morbidelli, A., Gladman, B., Petit, J.-M. 2000a, ``Understanding the distribution of Near-Earth Asteroids", Science, 288, pp. 2190-2194.

Bottke, W.~F., Morbidelli, A., Jedicke, R., et al. 2002, ``Debiased orbital and size distribution of the Near Earth Objects", Icarus, 156, pp. 399-433.

Bottke, W.~F., Rubincam, D.~F., Burns, J. ~A. 2000b, ``Dynamical evolution of main belt meteoroids: Numerical simulations incorporating planetary perturbations and Yarkovsky thermal forces", Icarus, 145, pp. 301-331.

Carruba, V., Huaman, M., Domingos, R.~C., Roig, F. 2013, ``Chaotic diffusion caused by close encounters with several massive asteroids II. The regions of (10) Hygiea, (2) Pallas, and (31) Euphrosyne", \aa, 550(A85), doi: 10.1051/0004-6361/201220448.

Carruba, V., Huaman, M., Douwens, S., Domingos, R.~C. 2012, ``Chaotic diffusion caused by close encounters with several massive asteroids. The (4) Vesta case", \aa, 543(A105), doi: 10.1051/0004-6361/201218908.

Delisle, J.-B., Laskar, J., 2012, ``Chaotic diffusion of the Vesta family induced by close encounters with massive asteroids", \aa, 540(A118), doi: 10.1051/0004-6361/201118339.

Greenstreet, S., Ngo, H., Gladman, B. 2012, ``The Orbital Distribution of Near-Earth Objects inside Earth's Orbit", Icarus, 217, pp. 355-366.

Machuca, J. F., Carruba, V. 2012, ``Secular dynamics and family identification among highly inclined asteroids in the Euphrosyne region", \mnras, 420, pp. 1779-1798.

Mainzer, A., Grav, T., Masiero, J. et al. 2012, ``Characterizing Subpopulations within the near-Earth Objects with NEOWISE: Preliminary Results", ApJ, 752(110), doi: 10.1088/0004-637X/752/2/110.

Milani, A., Gronchi, G.~F. 2009, Theory of Orbit Determination, Cambridge University Press.

Morbidelli, A., Bottke, W.~F., Froeschl\'{e}, C., Michel, P. 2002, ``Origin and evolution of Near Earth Asteroids", Asteroids III, University of Arizona Press, pp. 409-422.

}

\end{document}